\documentclass[sigconf]{acmart}
\usepackage{xcolor}
\usepackage{hyperref}
\usepackage{subcaption}
\usepackage{graphicx}
\usepackage{xspace}
\usepackage{listings}

\usepackage{tabularx}
\usepackage{multirow}

\newenvironment{change}{\color{black}}{\color{black}}

\definecolor{light-gray}{gray}{0.9}
\definecolor{dark-gray}{gray}{0.6}

\AtBeginDocument{}

\copyrightyear{2024}
\acmYear{2024}
\setcopyright{rightsretained}
\acmConference[ICPC '24]{32nd IEEE/ACM International Conference on Program Comprehension}{April 15--16, 2024}{Lisbon, Portugal}
\acmBooktitle{32nd IEEE/ACM International Conference on Program Comprehension (ICPC '24), April 15--16, 2024, Lisbon, Portugal}\acmDOI{10.1145/3643916.3644402}
\acmISBN{979-8-4007-0586-1/24/04}

\begin{document}

\title{Generating Java Methods: An Empirical Assessment of Four AI-Based Code Assistants}

\author{Vincenzo Corso}
\affiliation{\institution{University of Milano - Bicocca}
  \city{Milan}
  \country{Italy}
}
\email{v.corso3@campus.unimib.it}

\author{Leonardo Mariani}
\affiliation{\institution{University of Milano - Bicocca}
  \city{Milan}
  \country{Italy}
}
\email{leonardo.mariani@unimib.it}

\author{Daniela Micucci}
\affiliation{\institution{University of Milano - Bicocca}
  \city{Milan}
  \country{Italy}
}
\email{daniela.micucci@unimib.it}

\author{Oliviero Riganelli}
\affiliation{\institution{University of Milano - Bicocca}
  \city{Milan}
  \country{Italy}
}
\email{oliviero.riganelli@unimib.it}
\newcommand{\LEO}[1]{\textcolor{blue}{{\it [Leonardo says: #1]}}}
\newcommand{\DAN}[1]{\textcolor{red}{{\it [Daniela says: #1]}}}

\newcommand{\oli}[1]{\textcolor{red}{#1}}

\newcommand{\urlRepo}[0]{\url{https://osf.io/3mkqz/?view_only=ba3aed33c7b8488ca6aefdddd93a924c}\xspace}

\begin{abstract}
AI-based code assistants are promising tools that can facilitate and speed up code development. They exploit machine learning algorithms and natural language processing to interact with developers, suggesting code snippets (e.g., method implementations) that can be incorporated into projects. Recent studies empirically investigated the effectiveness of code assistants using simple exemplary problems (e.g., the re-implementation of well-known algorithms), which fail to capture the spectrum and nature of the tasks actually faced by developers. 

In this paper, we expand the knowledge in the area by comparatively assessing four popular AI-based code assistants, namely GitHub Copilot, Tabnine, ChatGPT, and Google Bard, with a dataset of 100 methods that we constructed from real-life open-source Java projects, considering a variety of cases for complexity and dependency from contextual elements. Results show that Copilot is often more accurate than other techniques, yet none of the assistants is completely subsumed by the rest of the approaches. Interestingly, the effectiveness of these solutions dramatically decreases when dealing with dependencies outside the boundaries of single classes.
\end{abstract}

\ccsdesc[500]{Software and its engineering~Integrated and visual development environments}
\keywords{AI-based code assistants, code completion, Copilot, ChatGPT, Tabnine, Bard, empirical study.}

\maketitle

\section{Introduction}
Artificial Intelligence (AI) has made outstanding progress in various domains, changing the way we interact with technology. In the realm of software development, AI-based code assistants have emerged as invaluable aids, empowering programmers to enhance their productivity and streamline the coding process \cite{raychev2014code,Allamanis:SurveyML:2018}. These tools employ advanced machine learning algorithms and natural language processing techniques to suggest code snippets, auto-complete syntax, and provide contextual guidance. As the complexity of software development continues to increase, the demand for effective code completion tools becomes crucial for enhancing developer productivity. 

For example, Copilot~\cite{Copilot2023} is a code assistant trained with code available in GitHub repositories to assist developers with code recommendations. It can generate a correct implementation of a method that calculates the greatest common divisor starting from a code comment as simple as ``\emph{Get the greatest common divisor between two numbers'}'. Similarly, ChatGPT~\cite{ChatGPT2023} is a text-based chatbot trained on a wide range of sources (e.g., books, articles, websites) to generate human-like responses on various topics including programming. It can recommend a correct implementation of a method that calculates the factorial of a number starting from a request like ``\emph{Write a method that calculates the factorial of a given number'}'.

Recent studies demonstrated that AI-based code assistants can indeed provide useful code snippets~\cite{Vaithilingam22,Yetistiren:PROMISE:22,Dakhel:JSS:2023}. For instance, Copilot has been useful with common programming problems~\cite{Yetistiren:PROMISE:22, Fagadau:Prompt:ICPC:2024} and has been observed to generate code of appropriate complexity when dealing with LeetCode~\cite{LeetCode2023} programming questions~\cite{Nguyen:MSR22}. These studies generated initial knowledge about the effectiveness of AI-based code assistants, but they also suffer from several key limitations. 

First, they are based on \emph{exemplary problems} that are quite different from the ones that developers face in reality. Algorithms to calculate the maximum in an array or delete a node in a linked list are normally available in libraries that developers can just reuse in their implementations, rather than implementing from scratch. In addition, sample code that solves these exemplary problems is largely available in online repositories and on the Web, and thus training a model to generate solutions to these problems is not particularly challenging. On the contrary, developers' code is often non-algorithmic. In fact, a substantial portion of projects' code is devoted to tasks, such as error handling, data validation and transformation, user interface logic, architecture, and event handling. 

Second, most of the existing studies consider the generation of code with \emph{no dependency} from the codebase, that is, the solution does not require invoking the other methods present in the project. While, in practice, the majority of methods have dependencies from other projects' methods, as also reported by our study.

Finally, existing studies focus on \emph{individual AI-based code assistants}, not disclosing findings about their relative effectiveness.

This paper addresses this gap by presenting a \emph{comparative study of four AI-based code assistants}, namely GitHub Copilot, Tabnine, ChatGPT, and Google Bard, involved in the generation of Java methods. The presented study considers \emph{programming tasks extracted from code changes present in GitHub}, thus representing a sample of the implementation tasks normally faced by developers. 
We investigate the quality of the generated code by considering functional correctness, but also \emph{complexity, efficiency, and size}. In particular, we exploit the unit test cases implemented by developers, jointly with manual code inspection, to assess the correctness of the generated code. Further, we exploit static and dynamic analysis tools to measure code complexity, efficiency, and size.

Finally, we investigate how AI-based code assistants can deal with \emph{increasingly larger portion of the context}. 
This is an important dimension since the code that is normally written is not standalone, but interacts with the rest of the codebase, and adapting the code generation routine to the specific context is a non-trivial challenge for code-generation tools. In our study, we distinguish between self-contained methods, methods with intra-class dependencies, and methods with external dependencies.  

The actual dataset we constructed for this investigation consists of 100 methods obtained from open-source projects. To mitigate the risk of using the AI-based code assistants to generate code that is already present in their training set, we limited the study to methods added \emph{after the most recent training of the compared tools}. This requirement is not explicitly considered in previous studies, but it is extremely important to prevent biases in the evaluation.

The results generate valuable insights about the strengths and limitations of AI-based code assistants. Copilot proved to be more effective than competing approaches, but interestingly each assistant generated at least a correct method implementation that the other three assistants failed to generate, suggesting collaboration between assistants as a promising research direction. On one hand, our results show that AI-based code assistants have still to be largely improved, especially when addressing inter-class dependencies. On the other hand, the generated code is sometimes even better than the code implemented by the developers (according to some dimensions), demonstrating how promising these tools are. 

In a nutshell, this paper contributes to the knowledge in the area by: (i) defining an experimental methodology and releasing\footnote{Experimental material to replicate our study is available at \urlRepo} the underlying dataset of 100 Java methods selected from open source projects in Github (see Section~\ref{sec:methodology}); (ii) reporting an experimental comparison of the Java methods generated by four state-of-the-art AI-based code assistants, based on the correctness, complexity, efficiency, size, and adherence of the generated code (see Section ~\ref{sec:results}); (iii) disclosing a set of findings that may pave the way to more work in the area (see Section~\ref{sec:findings}).

\section{AI-Based Code Assistants} \label{sec:AITools}

Code completion is a key feature in modern software development environments~\cite{Amann:SANER:2016,proksch2018enriched}. It suggests code and auto-completion options based on the context and syntax of the code being written. Solutions have evolved over the years, from simple keyword-based suggestions~\cite{little2007keyword} to more sophisticated tools that leverage machine learning and artificial intelligence techniques~\cite{GoogleBard2023,VisualStudioCode2023,Tabnine2023,ChatGPT2023,Copilot2023}.

In this paper, we consider four popular AI-based code assistants: Copilot~\cite{Copilot2023},  Tabnine~\cite{Tabnine2023}, ChatGPT~\cite{ChatGPT2023}, and Bard~\cite{GoogleBard2023}.

\emph{Copilot}~\cite{Copilot2023} is an AI-powered code completion tool developed by GitHub and OpenAI.  \begin{change}Copilot is powered by the OpenAI Codex model. Codex is a descendant of the GPT-3 architecture, fine-tuned specifically for code-related tasks.\end{change} The training process entails exposing the models to an extensive dataset comprising diverse code repositories across multiple programming languages. By learning from vast code repositories, Copilot can identify common code constructs, APIs, and libraries, finally generating context-aware code suggestions. Copilot works within the IDE by analyzing the context of the code being written, including variables, functions, and surrounding code, to provide code suggestions.

\emph{Tabnine}~\cite{Tabnine2023}, similarly to Copilot, is an AI-powered code completion tool that enhances the software development process by providing code suggestions within integrated development environments (IDEs). \begin{change}Although the details of the underlying model are not widely disclosed, also Tabnine's training involves the extensive analysis of diverse code repositories.\end{change}

\emph{ChatGPT}~\cite{ChatGPT2023}, \begin{change} based on the GPT-3 architecture,\end{change} is a powerful language model developed by OpenAI that can be utilized as a tool for generating code completions and suggestions. Compared to Copilot and Tabnine, ChatGPT excels at understanding natural language queries and providing detailed responses, while Tabnine and Copilot are specifically designed for code completion and have a deep understanding of programming languages. 

\emph{Bard}~\cite{GoogleBard2023}, as ChatGPT, is an AI assistant based on  
 \begin{change}the Pathways Language Model 2 (PaLM 2), which is a large language model (LLM) developed by Google AI.\end{change} Bard is trained on a large dataset of code from various open-source projects specifically to understand programming languages and to perform coding tasks.
A key difference between  Bard and ChatGPT is that Bard is trained specifically with code, while ChatGPT is trained with a more general dataset that covers a  wide range of topics (including programming). 

We decided to exploit these four tools to generate the implementation of Java methods for three reasons: (1) Java is a language well supported by all four tools; (2) Java code is largely present in public repositories, thus representing a good use case for AI-based code assistants, contrarily to other languages that may harm the capabilities of the tools due to the sometime yet limited resources available; and (3) Java is also well supported in terms of publicly available tools for measuring the quality of the generated code.

\section{Methodology} \label{sec:methodology}

In this paper, we assess AI-base code assistants by investigating five research questions:

\noindent \textbf{RQ1 - Is the code generated by AI-based code assistants \emph{correct}?} This RQ investigates the syntactic and semantic correctness of the generated code.

\noindent \textbf{RQ2 - What is the \emph{McCabe complexity} of the generated code?} This RQ investigates if AI-based code assistants are able to not only generate correct code but also produce code with a level of McCabe complexity similar to the code implemented by developers.

\noindent \textbf{RQ3 - How \emph{efficient} is the generated code?} This RQ investigates if the generated correct code is as efficient as the one implemented by developers.

\noindent \textbf{RQ4 - What is the \emph{size} of the generated code?} This RQ investigates if the generated correct code has a similar size to the one implemented by developers. 

\noindent \textbf{RQ5 - How \emph{far} is the generated code from the one implemented by developers?} This RQ studies the similarity of the code implemented by the developers to the code generated by the experimented tools, according to change-oriented static metrics. 

The rest of this section describes the methodology that we used to answer our five RQs using the four AI-driven code completion tools described in Section~\ref{sec:AITools}, reporting first how we constructed the dataset for the evaluation (Section~\ref{sec:dataset}), and then the metrics that we computed to answer each research question (Section~\ref{sec:assessment}).

\subsection{Dataset Construction} \label{sec:dataset}
In order to answer RQ1-5, we created a dataset of Java methods with Javadoc comments that can be passed to AI-driven code completion tools to obtain a possible implementation of the method. To select these methods, we referred to the criteria below.

\emph{Real-world methods}: The assessment must be based on methods that are part of real software systems, representing the actual problems that developers face when developing their systems. For this reason, we targeted methods that are part of open-source projects and avoided selecting any method that encodes simple didactic programming problems. 

\emph{Methods of different complexity levels}: Not all the methods that must be implemented have the same level of complexity. To sample the problem space considering a range of situations, we select methods of different complexity along two key dimensions. The first one is cyclomatic complexity, to guarantee that code with different levels of structural complexity is considered in the investigation. The second one is context dependency, to guarantee that code with various degrees of dependency from the rest of the code in the system is considered in the investigation.

\emph{Methods outside the scope of training}: Since AI-driven code completion tools are trained on the existing codebases, it might be that a method selected online to assess these tools has been also used in the training stage of the models used by these tools. To mitigate this threat, we only work with code added to public codebases after the last known date of training of these tools.

\begin{figure*}
\centering
\includegraphics[width=\textwidth]{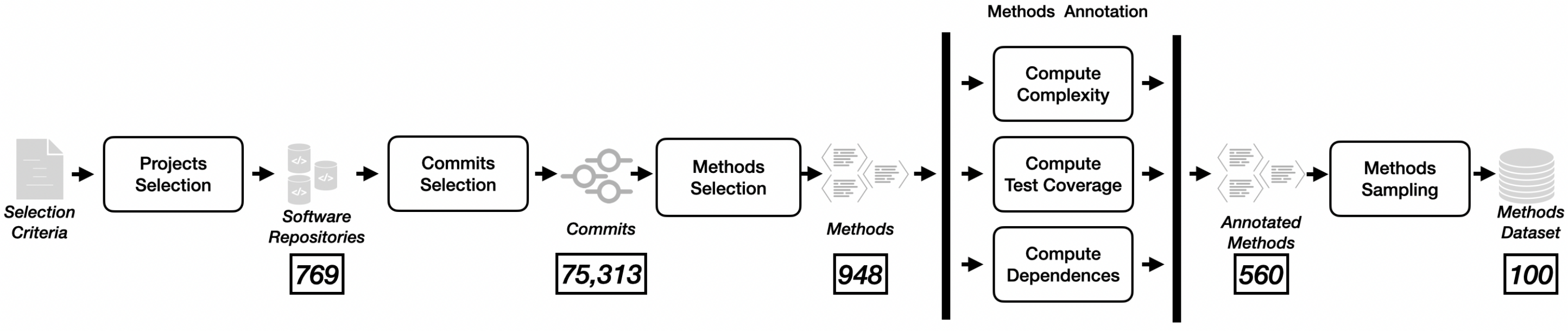}
\caption{Construction of the dataset.}
\label{fig:methodology}
\end{figure*}

These criteria map into the dataset construction methodology shown in Figure~\ref{fig:methodology}.

\textbf{Projects Selection}. We start by selecting the repositories, and thus the projects, where the methods will be taken from. We focus on the code present in GitHub, since it is the largest public repository of code and it implements APIs that can be used to automate the extraction process. We selected all the projects that satisfy the following constraints:

\noindent \emph{Maven Java projects}: We target Java projects (defined as having at least 80\% of their code implemented in Java), using Maven as a build tool to automate the various stages of the analysis. 

\noindent \emph{Recent commits}: To mitigate the issue of using code already used while training the compared models, we focus on projects with new code added between January 1st, 2023, and June 10th, 2023. In fact, although not all the tools always provide explicit information about the last training phase, the services we used have all been released in 2022. Repositories with no commits in this period are discarded. 

\noindent \emph{Well-ranked projects}: GitHub users can give positive feedback to projects by assigning stars. To focus the selection on high-quality projects that have been well received by the community, we select projects with at least $500$ stars. 

These selection criteria lead to the identification of $792$ project repositories. We manually checked these repositories to discard projects that are not about real software projects, such as repositories with a collection of patterns or example projects. We ended up with a list of $769$ project repositories.

\textbf{Commits Selection}. To identify recent commits, we select every commit in the interval between January 1st, 2023, and June 10th, 2023 that adds at least 5 lines of code. This returns 75,313 commits.

\textbf{Methods Selection}. 
To identify the relevant methods for our analysis, we only consider \emph{newly added} methods in the selected commits. We discard the methods shorter than 5 lines of code (header included), to exclude trivial code from our analysis (e.g., the implementation of getter methods), and test methods, since we focus on application methods.
We also discard methods without JavaDoc comments, which would not be suitable for this study. We use antlr (\url{https://www.antlr.org/}) to parse the methods added in the commits and automatically discover if they have an associated JavaDoc comment. In addition, we check for the presence of a test class for the classes where the methods are implemented. We look for test classes that follow the Maven naming convention, that is, given a class \texttt{Hello.java}, its test class could be named any of \texttt{HelloTest}, \texttt{HelloTests}, \texttt{HelloTestCase}, or \texttt{TestHello}. Methods with no tests could not be automatically validated for correctness in our study and are thus discarded. We finally discard methods in classes that are not part of the software product, that is, they are outside the scope of the \texttt{src/main} folder. To efficiently analyze such a large volume of projects, commits, and methods, we exploit git blobless cloning to extract the project metadata and download the actual content only when needed. After this step, we ended up with  948 methods eligible for our study.

\textbf{Methods Annotation}. In this stage, we both filter and annotate methods. Since we use the test suites available in the projects to assess the code, we focus on well-tested methods, that is, methods with at least 80\% statement coverage. We thus executed the test suites available on the project, computed the coverage, and discarded methods that were not thoroughly tested. 

To support the next step of the methodology, we analyze these methods to derive information about the dependencies and the cyclomatic complexity. We use DependencyFinder~\cite{DependencyFinder} to compute dependencies and classify methods in three categories: \emph{self-contained methods}, which are the methods with no dependency from the project's code, \emph{class-dependant methods}, which are the methods with at least a dependency from another method in the same class, and no dependencies with the rest of the code in the project, and \emph{external-dependant methods}, which are the methods with at least a dependency from another method in a different class of the project. 
We compute the McCabe cyclomatic complexity~\cite{McCabe76} of the methods with Understand~\cite{understand}. 

If the data about coverage, dependency, or complexity could not be computed, we discarded the method. We ended up with 560 methods annotated with information about test coverage, dependencies, and complexity suitable for our study. 

According to dependencies, the selected methods are distributed as follows: 53 self-contained methods, 101 class-dependant methods, and 406 external-dependant methods.  Note how the majority of the methods have dependencies with other methods in different classes of the project, confirming the intuition that code generation tools should be mainly validated with code that has dependencies from the context, and not with exemplary problems that do not need to exploit information from the context to be solved. According to complexity, the methods range from $1$ to $25$, with a median of $2$.

\textbf{Methods Sampling}.
To work with a dataset of a size manageable for the experiment, we extracted 100 methods from our selection of 560 annotated methods. We selected the methods to cover cases with different cyclomatic complexity and different degrees of dependency from the context. We also focus on recent thoroughly tested methods selected from a variety of highly popular projects.  
To fulfill these criteria, we applied the following procedure. 

We grouped the methods in each category (self-contained, class-dependant, and external-dependant) into buckets of equal cyclomatic complexity. For instance, the self-contained methods are split into seven buckets, the first bucket includes eight methods of cyclomatic complexity $1$, the second bucket includes $14$ methods of cyclomatic complexity $2$, and so on.

We sorted the buckets in each category in ascending order of cyclomatic complexity. We sorted the methods within each bucket in descending order based on the number of stars of the project they belong to (assigning higher positions to methods from well-ranked projects). In case of a tie in the number of stars, we sorted them based on the statement coverage provided by the test suites (assigning higher positions to the better tested methods). Finally, if the coverage is also the same, we sorted them based on the commit date of the commit they belong to (assigning higher positions to the most recent methods).

We selected a different number of methods from each category to reflect the non-uniform size of the three categories: 25 self-contained methods, 30 class-dependant methods, and 45 external-dependant methods. 

\begin{figure}
\centering
  \includegraphics[width=\columnwidth]{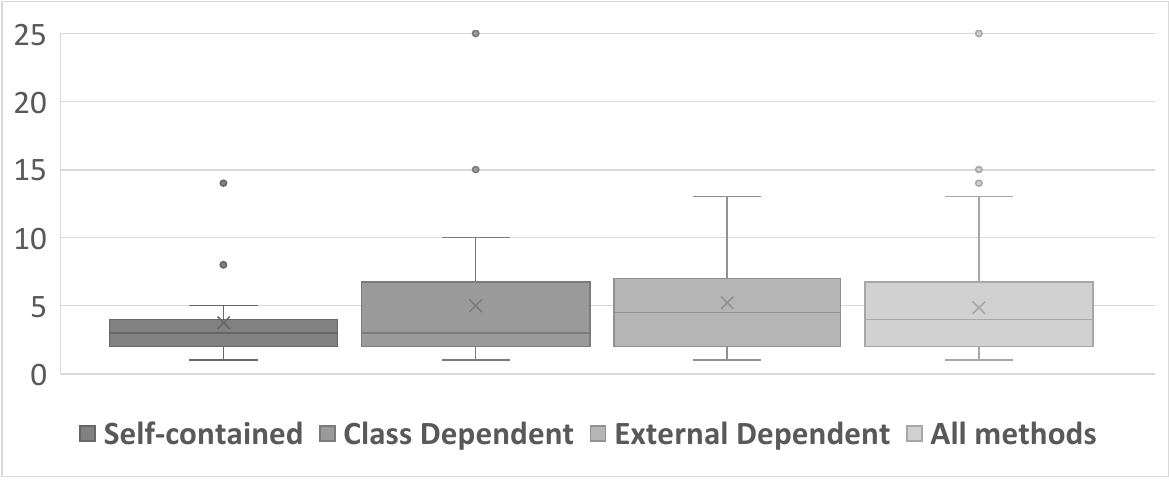}
\caption{Distribution of methods according to complexity.}
\label{fig:afterSampling}
\end{figure}

\begin{lstlisting}[basicstyle=\small\ttfamily,breaklines=true,columns=fullflexible, frame=single, label=code, caption={An example selected method from our dataset.}]

/**
 * If the toleration value is an empty string, set it to
 * null. That solves an issue when built STS contains a 
 * field with an empty property value. K8s is removing  
 * properties like this, and thus we cannot fetch an equal 
 * STS which was created with (some) empty value.
 *
 * @param tolerations   Tolerations list to check whether 
 * toleration value is an empty string and eventually
 * replace it by null
 *
 * @return              List of tolerations with fixed empty strings
*/
public static List<Toleration> removeEmptyValuesFromTolerations (List<Toleration> tolerations) {
    if (tolerations != null) {
        tolerations.stream().filter(toleration -> toleration.getValue() != null && toleration.getValue().isEmpty()).forEach(emptyValTol -> emptyValTol.setValue(null));
        return tolerations;
    } else {
        return null;
    }
}    
\end{lstlisting}

We performed the selection of the methods used in the study from each category as follows. We start by selecting one method per bucket, to cover the various complexity cases. When the number of methods to be extracted is smaller than the number of available buckets, we extract the remaining methods from equidistant buckets, to still consider the full distribution of values. We use linspace in the numpy library~\cite{linspace} to select the buckets. 
For example, if there are 3 methods left to be extracted and there are 5 buckets with selectable methods, the 3 methods will be extracted from the first, third, and fifth buckets.
When extracting a method from a bucket, we select the first one (according to the sorting strategy described above) that belongs to a project from which no other method has been selected yet, to ensure a diversity of cases in our selection. If all the methods in the bucket belong to projects that have already been considered, then the method at the top of the bucket is extracted. 

Figure~\ref{fig:afterSampling} shows the complexity of the 100 selected methods \begin{change} from 52 distinct projects. Listing~\ref{code} shows an example of a selected method. The method takes a list of tolerations as an input and returns a list of tolerations with empty values replaced with \texttt{null}.
\end{change}The 100 selected methods are reported in our online repository: \urlRepo.

\subsection{Assessment of the Generated Code} \label{sec:assessment}
To answer the five research questions, we executed Copilot, Tabnine, ChatGPT, and Bard on the methods present in our dataset. 
In the case of Copilot and Tabnine, we installed their plugins (versions 1.2.8.2631 and  1.0.15, respectively) in the IntelliJ Idea editor (version 2023.1). For each method in the dataset, we opened in IntelliJ Idea the class containing the method to be generated, whose implementation was already deleted upfront, maintaining only the comment and the signature. In the case of ChatGPT and Bard, for each method in the dataset, we deleted its implementation from the enclosing class, again preserving the comment and the signature. We then entered the modified class and the prompt
{\scriptsize
\texttt{"Generate the implementation of <signature with visibility modifier and return type>}"}
 into the service interface. By construction, ChatGPT and Bard access to a restricted amount of context (i.e., text) compared to Copilot and Tabnine. We handled these cases by removing from the bottom of the class the minimum number of methods necessary to fit into the allowed length, making sure to not remove any method needed in the code to be generated. Despite this limitation, our results show that they have been both able to generate correct code with external dependencies. We considered the first suggestion provided by each tool. We performed the study in June 2023.

To answer RQ1, we investigated both the syntactic and semantic correctness of the generated code. We first determined the \emph{invalid methods}, that is, the methods that fail to compile because of syntactic problems (e.g., misplaced tokes or missing return statements in non-void methods). We then launched the test cases available with the project to determine the \emph{semantically incorrect methods}, that is, the methods that fail with at least one of the available test cases. The remaining methods are the \emph{plausibly correct methods}, which are the methods that pass the available test cases. To determine the actual \emph{correct} methods, we manually inspected all the plausibly correct methods and identified the subset of the correct methods that have the same semantic behavior as the original methods implemented by the developers according to human judgment. 
The inspection was performed by two authors of this paper who independently assessed the generated code. Out of the 124 plausibly correct methods inspected, the two inspectors agreed on 105 classifications. This corresponds to a substantial agreement according to Cohen's kappa ($k=0.69$). The conflicting classifications have been resolved in a meeting, with the two inspectors finally reaching a consensus on all the inspected methods. As a rule of thumb, we considered correct any method that is semantically equivalent to the original method or differs only for defensive checks included in the method (e.g., checking for parameters being different than \texttt{null}). Note that the correct methods are a subset of the plausibly correct methods.

 To answer RQ2, we compared the code complexity of the reference implementation in the dataset to the code generated by the studied tools. We performed the comparison only for the correct methods since there is no point in assessing the complexity of incorrect code that is not performing the same computation as the original code. 

To answer RQ3, we executed every available test case for the method under analysis 30 times on both the reference code and the generated code, to determine if specific choices made by the compared tools may have an impact on test execution time. We used a machine equipped with a CPU AMD Ryzen 5 3500u, 8GB RAM, and Ubuntu 22.04.3LTS for the experiments. The used version of the JVM is not unique and it depends on the requirements of the specific project that is tested. We used the maven-surefire-plugin to collect data~\cite{SureFirePlugin2023}.

Note that although efficiency correlates with code complexity, inefficient code might also result from different choices in the used API methods. For example, in one case the developer used the \texttt{Scanner} class to transform a string into a number, while the automatically generated code directly uses the static \texttt{parseInt} method of the \texttt{Integer} class, thus avoiding the instantiation of an object and the invocation of multiple methods. We run the comparison for the tests that exercise the correctly generated methods that are not identical to the original method implemented by the developers.

To answer RQ4, we calculate the difference in lines of code (LOC) between the generated code and the code implemented by developers. We utilized Understand~\cite{understand} to compute this metric.

To answer RQ5, we compute both the normalized Levenshtein similarity and the CodeBLEU metrics~\cite{CodeBLEU}. The former metric captures the amount of syntactic changes needed to obtain the original method from the generated one. The latter metric is a metric that combines information about syntactic correctness, grammar correctness, and logic correctness to accurately measure the distance between two code fragments, implicitly considering also the code style. For both metrics, the higher the value is, the more similar the compared code snippets are. 
We measure normalized Levenshtein similarity using the Levenshtein distance function\footnote{The returned distance is normalized and subtracted from 1 to obtain the normalized Levenshtein similarity.} in the Levenshtein Python library~\cite{Levenshteinsw} and CodeBLEU with CodeXGLUE, a tool developed by Microsoft and available on GitHub~\cite{CodeXGLUE}.

\section{Results} \label{sec:results}
This section presents the findings and analysis of the research questions about the use of AI-based code assistants.

\subsection{RQ1: Is the code generated by AI-based code assistants correct?}

\begin{table}
\caption{Results for RQ1.}
\resizebox{\columnwidth}{!}{
\begin{tabular}{llcccc}
 & & \textbf{Copilot} & \textbf{Tabnine} & \textbf{ChatGPT} & \textbf{Bard} \\ \toprule 
\multirow{4}{*}{\parbox{1.4cm}{Self-contained}} & \textit{Correct}   & \textbf{10 (50\%)}               & 2 (10\%)               & 6 (30\%)               & 5  (25\%)           \\
         & \textit{Plausible} & 14               & 5                & 10               & 7             \\
& \textit{Incorrect}  & 6                & 13                & 10                & 12             \\
		& \textit{Invalid}   & 0                & 2                & 0                & 1             \\ \midrule 
\multirow{4}{*}{\parbox{1.4cm}{Class-dependent}} 	& \textit{Correct}   & \textbf{14 (50\%) }             & 5 (18\%)              & 11 (39\%)              & 8  (29\%)           \\
					& \textit{Plausible} & 18               & 6                & 13               & 10            \\
& \textit{Incorrect}  & 6                & 10                & 11                & 9             \\
			               & \textit{Invalid}   & 4                & 12               & 4                & 9            \\ \midrule
\multirow{4}{*}{\parbox{1.4cm}{External-dependent}}  	& \textit{Correct}   & \textbf{8 (15\%) }              & 6 (11\%)               & 6 (11\%)               & 2  (4\%)           \\
		& \textit{Plausible} & 15               & 9                & 11                & 6             \\
& \textit{Incorrect}  & 14                & 18                & 16               & 16            \\
		& \textit{Invalid}   & 23               & 25               & 25               & 29            \\ \midrule \midrule
\multirow{4}{*}{\bf{Total}} 	& \textit{Correct}   & \textbf{32 (32\%)}              & 13  (13\%)             & 23 (23\%)             & 15 (15\%)           \\
               & \textit{Plausible} & 47               & 20               & 34               & 23            \\
& \textit{Incorrect}  & 26               & 41               & 37               & 37            \\
               & \textit{Invalid}   & 27               & 39               & 29               & 40            \\ \bottomrule
\end{tabular}
}
	\label{tab:correctness_results}
\end{table}

\begin{change}
Table~\ref{tab:correctness_results} provides an overview of the methods generated by Copilot, Tabnine, ChatGPT, and Bard, classified according to the four categories defined in Section~\ref{sec:assessment}: \emph{invalid}, \emph{incorrect}, \emph{plausible}, and \emph{correct} methods\footnote{Note that the correct methods are a subset of the plausible methods, and thus the total number of methods analyzed corresponds to the sum of the invalid, incorrect, and plausible methods.}. 

For example, Listing~\ref{codeCopilot} shows the method implementation generated by Copilot for the case shown in Listing~\ref{code}. This is the case of a plausibly correct method, since it passes all the test cases, but it is not a correct method because it misses to implement a \texttt{null} check on the input parameter, which causes a null pointer exception when \texttt{null} input values are entered. In this same case,  Bard generates an \emph{incorrect} method due to a hallucination error, since it includes an invocation to a method \texttt{value()} of the \texttt{Toleration} class, which does not exist in reality.

\begin{lstlisting}[float,basicstyle=\small\ttfamily,breaklines=true,frame=single,columns=fullflexible,label=codeCopilot, caption={Method generated by Copilot.}]

public static List<Toleration> removeEmptyValuesFromTolerations (List<Toleration> tolerations) {
    List<Toleration> result = new ArrayList<>();
    for (Toleration toleration : tolerations) {
        if (toleration.getValue() != null && toleration.getValue().equals("")) {
            result.add(new TolerationBuilder(toleration).withValue(null).build());
        } else {
            result.add(toleration);
        }
    }
    return result;
}    
\end{lstlisting}
\end{change}

Approach-wise Copilot generated the highest number of plausible and correct method implementations, and consequently the least number of incorrect and invalid methods. In particular, it generated 32\% correct and 47\% plausible implementations overall. ChatGPT achieved good results, although being significantly worse than Co-pilot, producing 23\% correct methods and 34\% plausible methods. Google Bard and Tabnine were the least effective, generating 15\% and 13\% correct methods respectively. 

Interestingly, the relative effectiveness of the four AI-based code assistants (Copilot then ChatGPT then Bard, and finally Tabnine, when ordered by the number of correct methods generated) is consistent across method categories, with the only exception of Tabnine performing better than Google Bard for external dependent methods, probably due to its capability to access a broader context.

\begin{figure}
\centering
  \includegraphics[width=0.7\columnwidth]{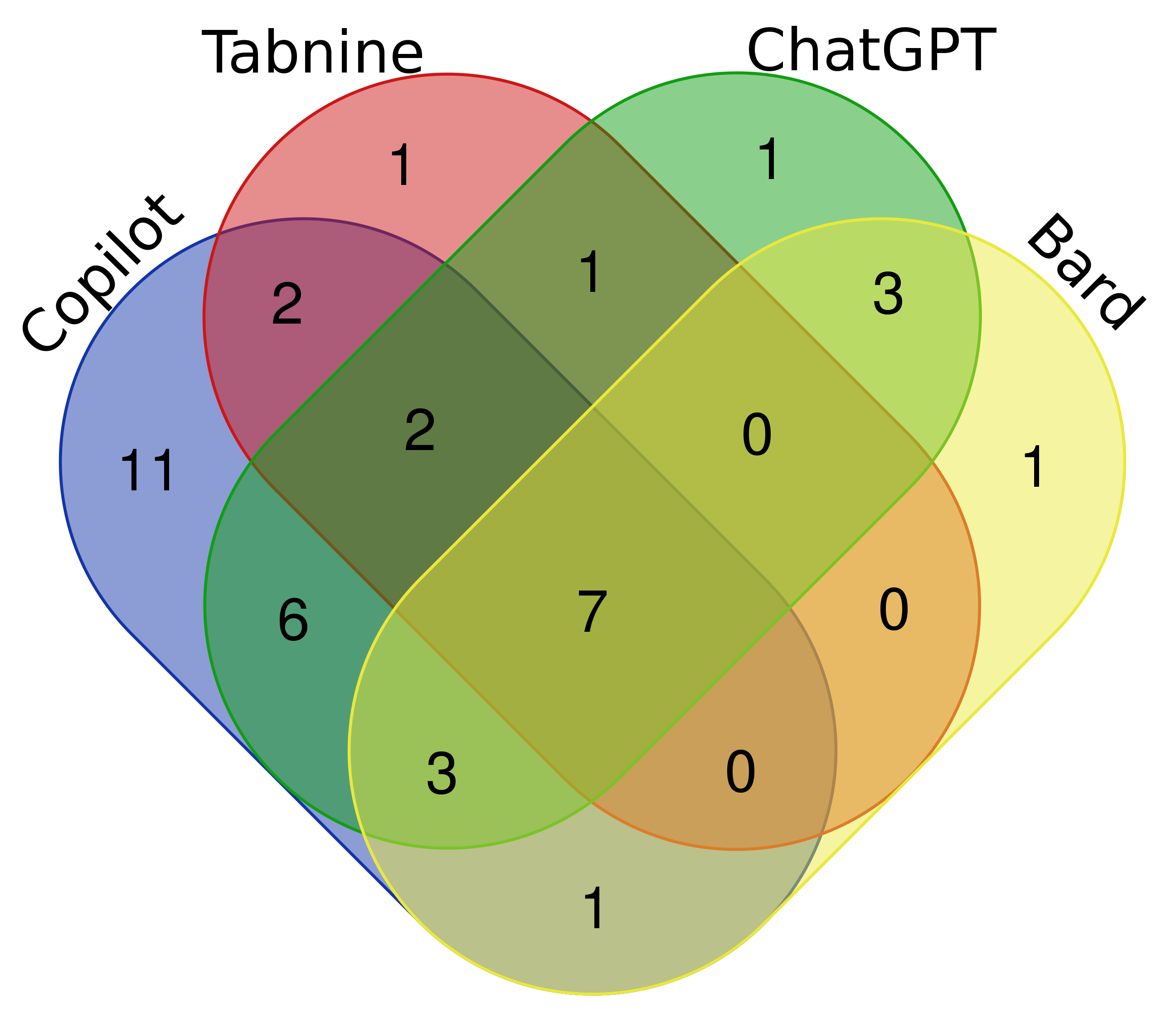}
  \Description{A}
  \caption{Generated methods that are correct.}
  \label{fig:correctness_results}
\end{figure}

Although the four approaches generated plausible and correct methods at different rates, none of them is entirely subsumed by the other approaches, even if considered altogether. Figure~\ref{fig:correctness_results} shows a Venn diagram of the correct methods generated by each technique. Notably, every technique generated methods that the other approaches were unable to generate, for a total of 39 correct methods collectively generated by the four approaches.  Copilot generated the highest number of correct methods not generated by other approaches (11 methods). We inspected these methods looking for common patterns and, although there was not a single common optimization, in 4 cases out of 11, Copilot better handled loops than competing techniques. 

Surprisingly, ChatGPT, which is the second best ranked approach, generated only one correct method that the other approaches could not generate. That method required the manipulation of strings using methods defined in external classes, suggesting that ChatGPT is the only assistant with knowledge about these external classes.

Tabnine and Bard, although performing worse than Copilot and ChatGPT, implemented one correct method that the other approaches failed to generate. Tabnine generated code identical to the original method, suggesting that very similar code might be present in its learning codebase. Bard generated code that uses the shift operator and that replaces library calls with alternative methods, which could not be generated by the other techniques.

It is interesting to report how ChatGPT and Bard sometimes generate defensive method implementations that check for the correctness of the parameters (e.g., checking for \texttt{null} values) before performing any computation, even if not present in the developers' code. Although these checks are not apparently needed, they are often good to have, since they reduce the risk of introducing regression errors during software evolution. Further, ChatGPT and Bard have been able twice to generate correct implementations without code smells that were in turn present in the developers' code. In both cases, the code smell consisted of failing to reuse a method already present in the class, introducing redundant code.

\emph{In a nutshell, although Copilot has been the most effective solution for the generation of method implementations, its capabilities do not subsume the capabilities of the other assistants. Each assistant generated at least a correct method not generated by the others, suggesting that studying how to combine the recommendations of multiple assistants, independently trained with different data sets, is a promising research direction.}

The analysis of method categories revealed that the external context plays a big role in the challenge of generating code. The effectiveness of the four tools is similar when addressing the generation of self-contained or class-dependent methods. For instance, Copilot and ChatGPT generated 50\% and 30\% correct self-contained methods and 50\% and 39\% class-dependent methods. On the contrary, the effectiveness of all the techniques drops dramatically when dealing with external dependent methods. For instance, Copilot, ChaptGPT, Tabnine, and Bard generated only 15\%, 11\%, 11\%, and 4\% correct external dependent methods. While this drop could be expected for ChatGPT and Bard, which have no access to external context, this result is surprising for Copilot and Tabnine.

\emph{The results obtained for RQ1 demonstrate there is significant space for improvement for all the techniques in all method categories. In fact, a non trivial portion of the generated code is invalid or incorrect (53\% of the methods are either invalid or incorrect for the best performing AI-based code assistant). The need for more sophisticated solutions is even more evident when the code to be generated depends on other code elements present in the project. In fact, the best performing technique only achieved 16\% correctness with external dependent code.}

 \subsection{RQ2: What is the McCabe complexity of the generated code?}
\begin{change} Code complexity is often indicative of code quality. Highly complex code can be more error-prone, and harder to understand, and maintain. An effective AI-driven code completion tool should be able to generate code that is not overly complex, promoting better code quality.\end{change} To assess the complexity of the generated code, we compared the McCabe cyclomatic complexity~\cite{McCabe76}  of the automatically generated code to the complexity of the code implemented by the developers. 

\begin{figure}
\centering
    \begin{subfigure}[t]{\columnwidth}
    \centering
    \includegraphics[width=\columnwidth]{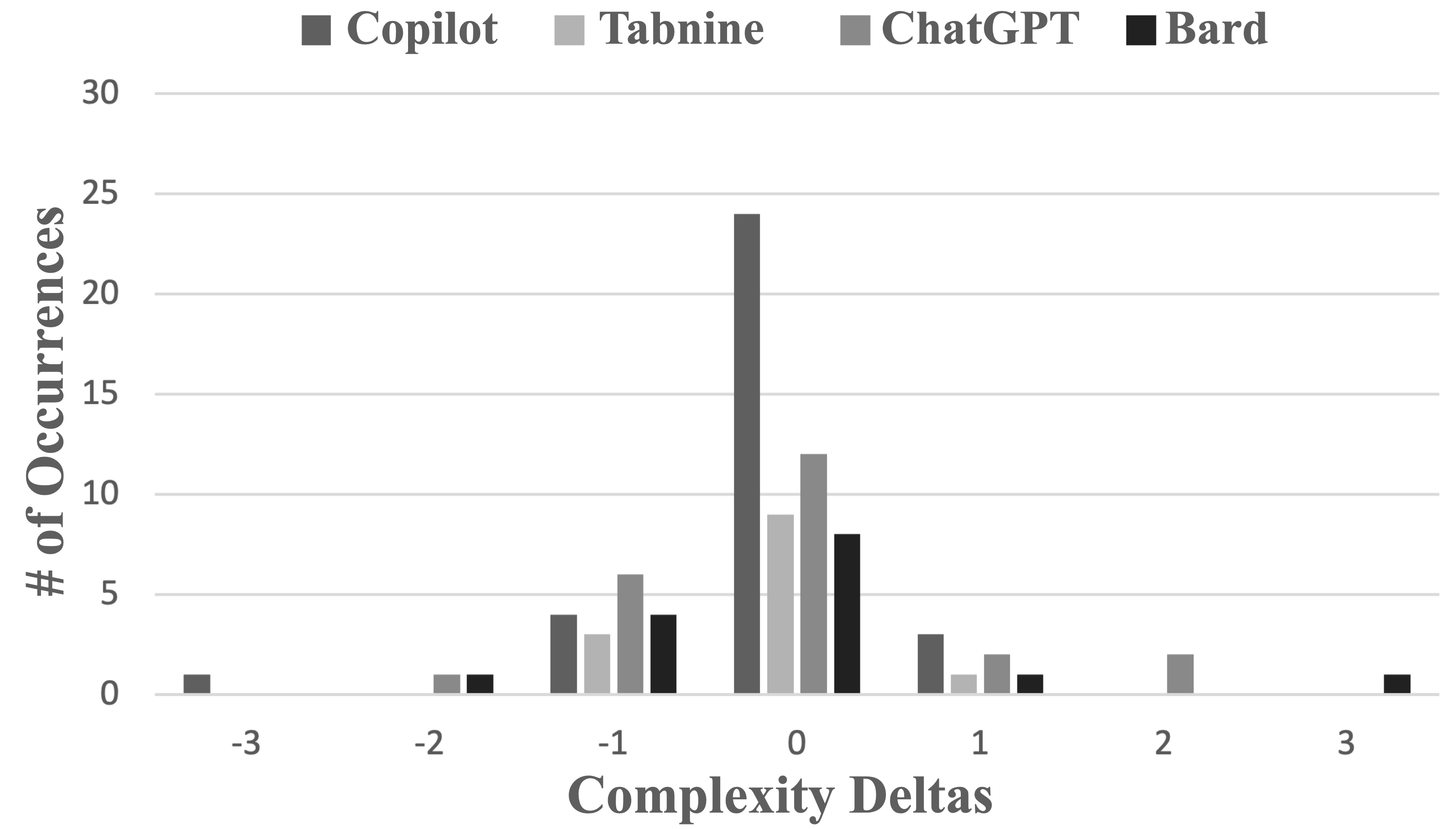}

    \caption{Methods generated correctly by all techniques} \label{fig:CCCommonMethods}
\end{subfigure}\hfill
\begin{subfigure}[t]{\columnwidth}
    \centering
    \includegraphics[width=\columnwidth]{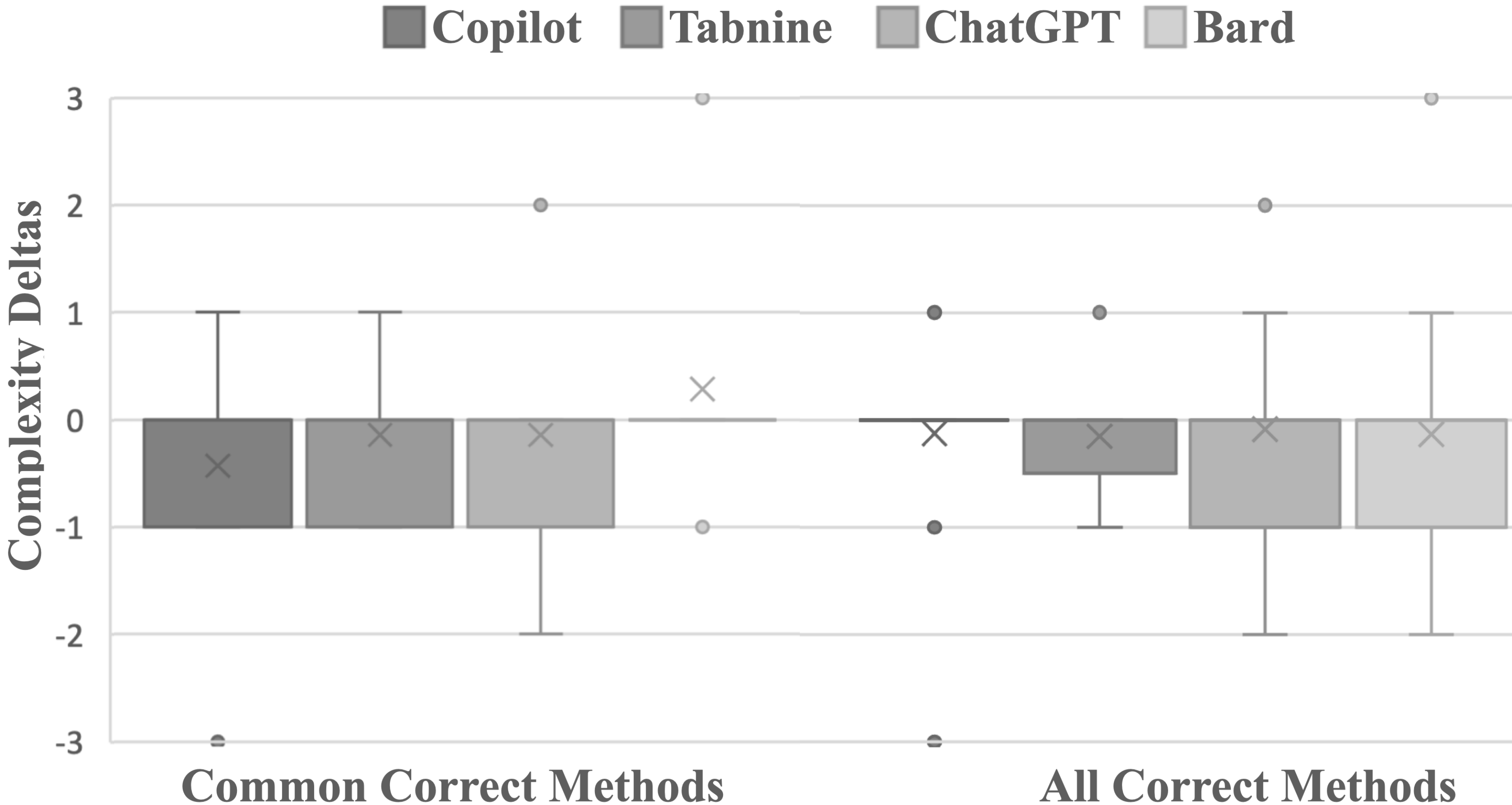}

    \caption{Complexity deltas for all correct and common correct methods.} \label{fig:CCallcorrectMethods}
\end{subfigure}\hfill

\caption{Cyclomatic Complexity Results.}
\label{fig:CC}
\vspace{-0.2cm}
\end{figure}

Figure~\ref{fig:CCCommonMethods} shows, for each tool, the number of correctly generated methods with a given \emph{complexity delta}, which is the difference between the cyclomatic complexity of the generated code and the cyclomatic complexity of the original code implemented by the developers. Positive, zero, and negative values indicate generated code with higher, same, and smaller complexity, respectively. For instance, the bars on -1 indicate the number of correctly generated methods with a cyclomatic complexity of $1$ point smaller than the code implemented by the developers. Figure~\ref{fig:CCallcorrectMethods} shows the complexity delta of the methods generated by each technique, considering both all the 40 correct methods, and only the 7 methods correctly generated by all the techniques\begin{change}, the latter being called common correct methods.\end{change}  

Overall, the four approaches, not surprisingly, tend to generate code of the same complexity as the code implemented by the developers. In some cases, the resulting code might have slightly higher complexity, between 1 and 3 in our experiments. \begin{change}This can often be attributed to the explicit usage of \texttt{if} conditions, as opposed to invoking methods that encapsulate checks. For instance, consider the \texttt{stream()} method in Listing~\ref{code}, which is a functional programming construct that does not introduce additional execution paths. In contrast, the code generated by Copilot in Listing~\ref{codeCopilot} employs an explicit \texttt{if}  condition.\end{change}

Interestingly, the approaches can sometimes procuce code with smaller complexity. For instance, the generated code sometimes uses simpler control flows to check for errors, or exploit lambda expressions, which decrease the cyclomatic complexity of the code.

When comparing the complexity of the code generated by the four approaches, they all behave similarly (pairwise differences among approaches are all non-significant according to the Mann-Whitney test with a significance level of 0.05). Copilot and Bard have been the only ones achieving a -3 and a +3 in complexity delta. The code with low complexity is due to the use of lambda expressions. While the unnecessarily complex code generated by Bard is due to the presence of defensive checks and the missing usage of the method \texttt{computeIfAbsent} of the \texttt{Map} interface. 

\emph{We can thus conclude that the complexity of the generated code often resembles the complexity of the code implemented by the developers, with some positive (simple code) and negative (unnecessarily complex code) exceptions. The four approaches reveal minimal differences in terms of complexity delta.}

\subsection{RQ3: How efficient is the generated code?}

\begin{table}[ht]
\caption{Performance of correct methods.} \label{tab:effoiciency} \resizebox{\columnwidth}{!}{
\begin{tabular}{lcccccc}
\toprule
	& \multicolumn{3}{c}{\textbf{All correct methods}} & \phantom{a} & \multicolumn{2}{c}{\textbf{Common Correct Methods}}\\
	\cmidrule{2-4} \cmidrule{6-7} 
         & \textbf{No} &\textit{\textbf{\# of slower }} & \textit{\textbf{\# of faster}} &&\textit{\textbf{\# of slower }}& \textit{\textbf{\# of faster}}\\
         & \textbf{Diff} &\textit{\textbf{methods}} & \textit{\textbf{methods}} && \textit{\textbf{methods}} & \textit{\textbf{methods}} \\ \midrule

Copilot	& 25 (78\%) & 4 (12\%)   & 3 (10\%)   && 0/7 & 1/7                    \\ 
Tabnine	& 9 (69\%) &0 (0\%)   & 4 (31\%)      && 0/7 & 3/7                                 \\
ChatGPT	&15 (65\%) & 3 (13\%) & 5 (22\%)     && 0/7 & 3/7                       \\ 
Bard		&9 (56\%) & 1  (6\%)    & 6 (38\%)      && 0/7 & 3/7                               \\ 
\bottomrule
\end{tabular}  
} 
\end{table}

To determine if the efficiency of a generated method significantly differs from the efficiency of the original method, we run the Mann-Whitney test with a significance level of 0.05 between the two sets of samples collected by running the same test on the two methods. If the test reveals significant differences, the tested method is reported. Table~\ref{tab:effoiciency} shows the number of methods that report no significant differences (Column \emph{No Diff}), and the methods with significant differences (i.e., with at least a test that exercises the methods showing significant differences in times between them), distinguishing between those with higher mean than the original method (Column \emph{\# of slower methods}) and those with lower mean than the original method (Column \emph{\# of faster methods}). We report numbers for all the correctly generated methods and specifically for those methods that are common among the four approaches.

All the techniques generated mostly methods that are either faster or have no significant differences compared to the code implemented by the developers (between 87\% and 100\% of the cases). This suggests that AI-based code assistants usually produce efficient code, as long as the code to be generated is not critical for performance (studying how the generators perform with code that requires highly efficient algorithms is out of the scope of this study).  

We notice small differences between the four assistants, with two of them producing more often than the other slower methods. The first one is Copilot, which generated the highest number of slower methods and the fewer faster methods, also when restricted to the seven correct methods generated by all the approaches. \begin{change}This is also the case of Listing \ref{codeCopilot}, where Copilot uses a loop instead of the highly optimized \texttt{stream()} method to filter the list of tolerances, as done in the developers' code shown in Listing~\ref{code}.\end{change} The second one is ChatGPT which generated three slower methods.

We inspected the slower methods, and those are mainly due to suboptimal data type choices and unnecessary operations (three occurrences),  and inefficient control flow and redundant method calls (five occurrences). We also inspected the faster methods, and those are mainly due to optimized memory allocation and use of efficient data access methods (two occurrences), avoiding unnecessary operations (seven occurrences), using functional methods and optimizing loops (four occurrences), and minimizing the number of method calls and conditionals (five occurrences).  

In general, the improvements are often small, with some outliers mainly due to the time saved by skipping simple operations. Examining the change percentages, we noted that Copilot exhibited considerable fluctuations, ranging from +80\% to -80\%. 
This wide range of change percentages highlights the need to understand the performance characteristics of proposed solutions when these are to be included in an application with stringent constraints on execution time. Contrarily, Bard, Tabnine, and ChatGPT produced code with small differences in the relative performance, with values generally ranging between -38\% and -13\%. In all the cases, absolute time differences ranged between -2.5s and 0.5s. These differences are unlikely to affect the actual performance of an application, unless they are spread in the code or introduced in specific methods.

\emph{Overall, this result suggests that the generated code has usually an efficiency that is either aligned with the developers' code or faster, for a number of cases between 87\% (ChatGPT) to 100\% (Tabnine). On the other hand, sometime code may include inefficiencies, with Copilot and ChatGPT being responsible for the majority of these cases.} \subsection{RQ4: What is the size of the generated code?}

\begin{figure}
\centering
  \includegraphics[width=\columnwidth]{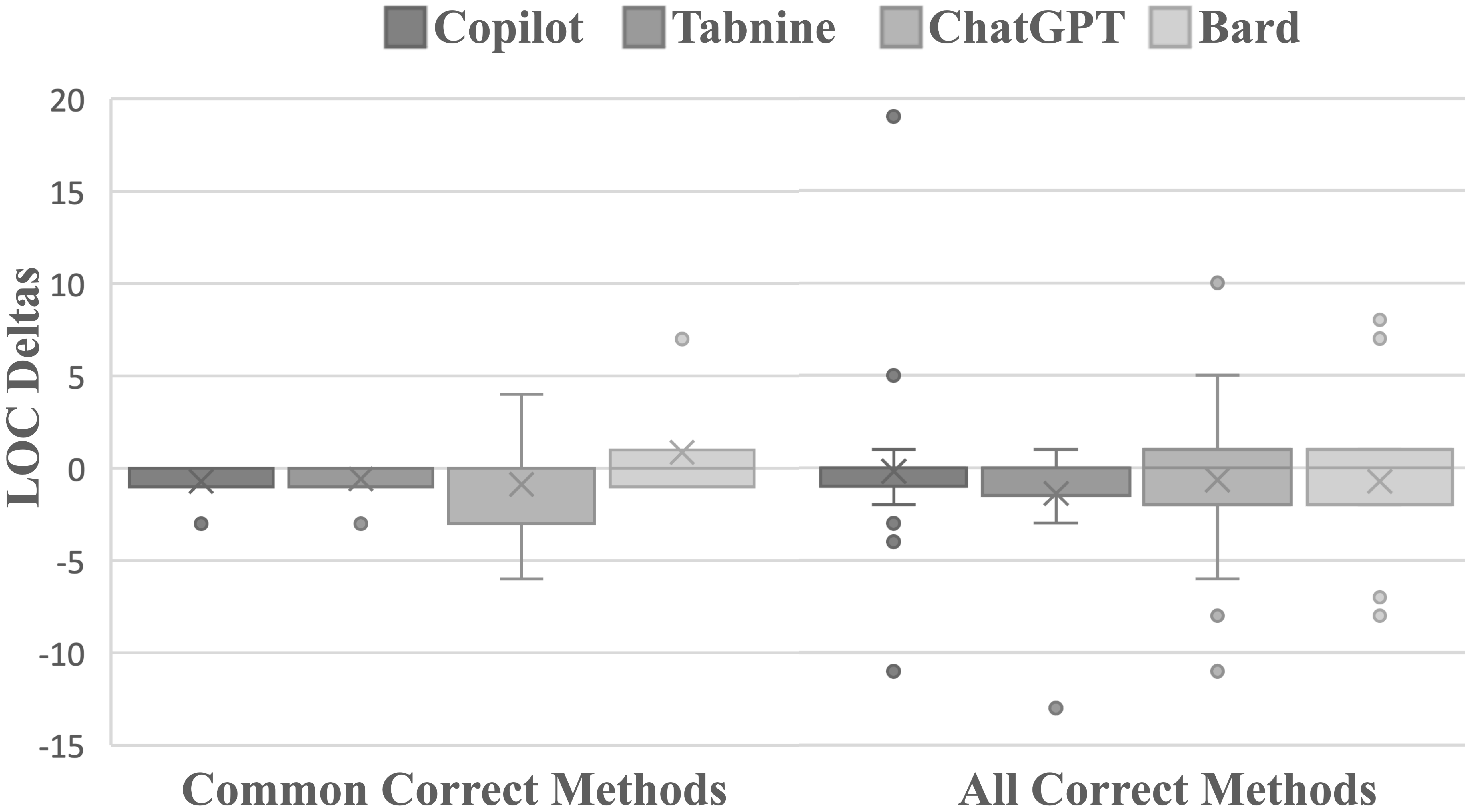}  
  \caption{Lines of Code Deltas.}
  \label{fig:loc}
  \vspace{-0.2cm}
\end{figure}

We studied the size of the generated code considering the deltas of LOCs, compared to the code implemented by the developers. Figure~\ref{fig:loc} shows LOCs deltas for all the correct methods and the methods in common among the four techniques. Indeed, the four approaches all generated code of a similar size to the code implemented by the developers, with no remarkable differences among techniques (confirmed by the lack of statistical differences according to the Mann-Whitney test with a significance level of 0.05).

In one case, Copilot generated code with many more lines compared to the original solution due to the repeated usage of \texttt{if} conditions instead of the ternary operator \texttt{? :}.

\emph{Based on the collected evidence, LOCs are similar for the generated and developers' code.}

 \subsection{RQ5: How far is the generated code from the one implemented by developers?}

\begin{figure}
\centering
\includegraphics[width=\columnwidth]{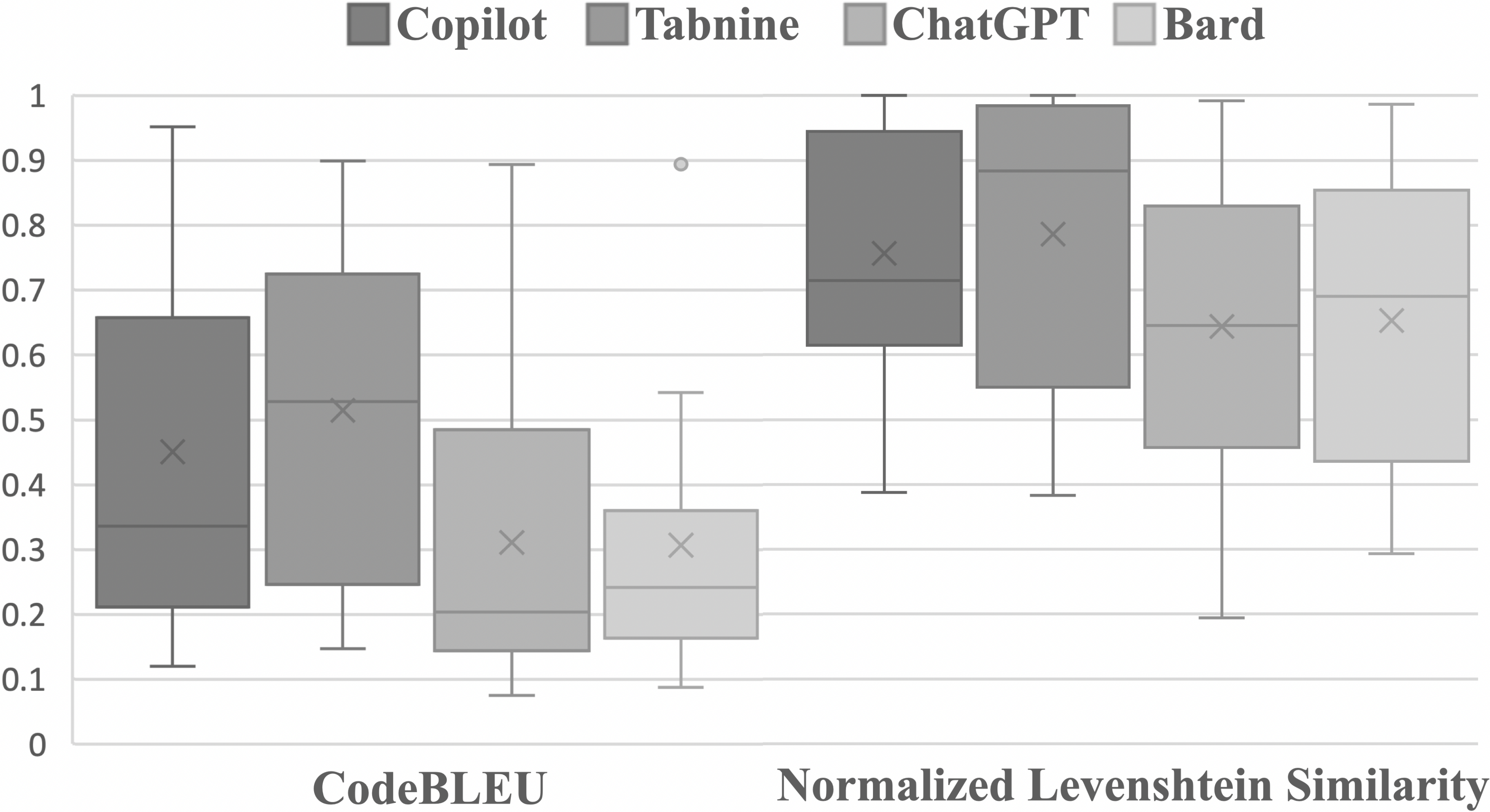}
  \caption{Similarity of correct methods.}
  \label{fig:similarityCorrect}
\end{figure}

\begin{figure} 
\centering
\includegraphics[width=\columnwidth]{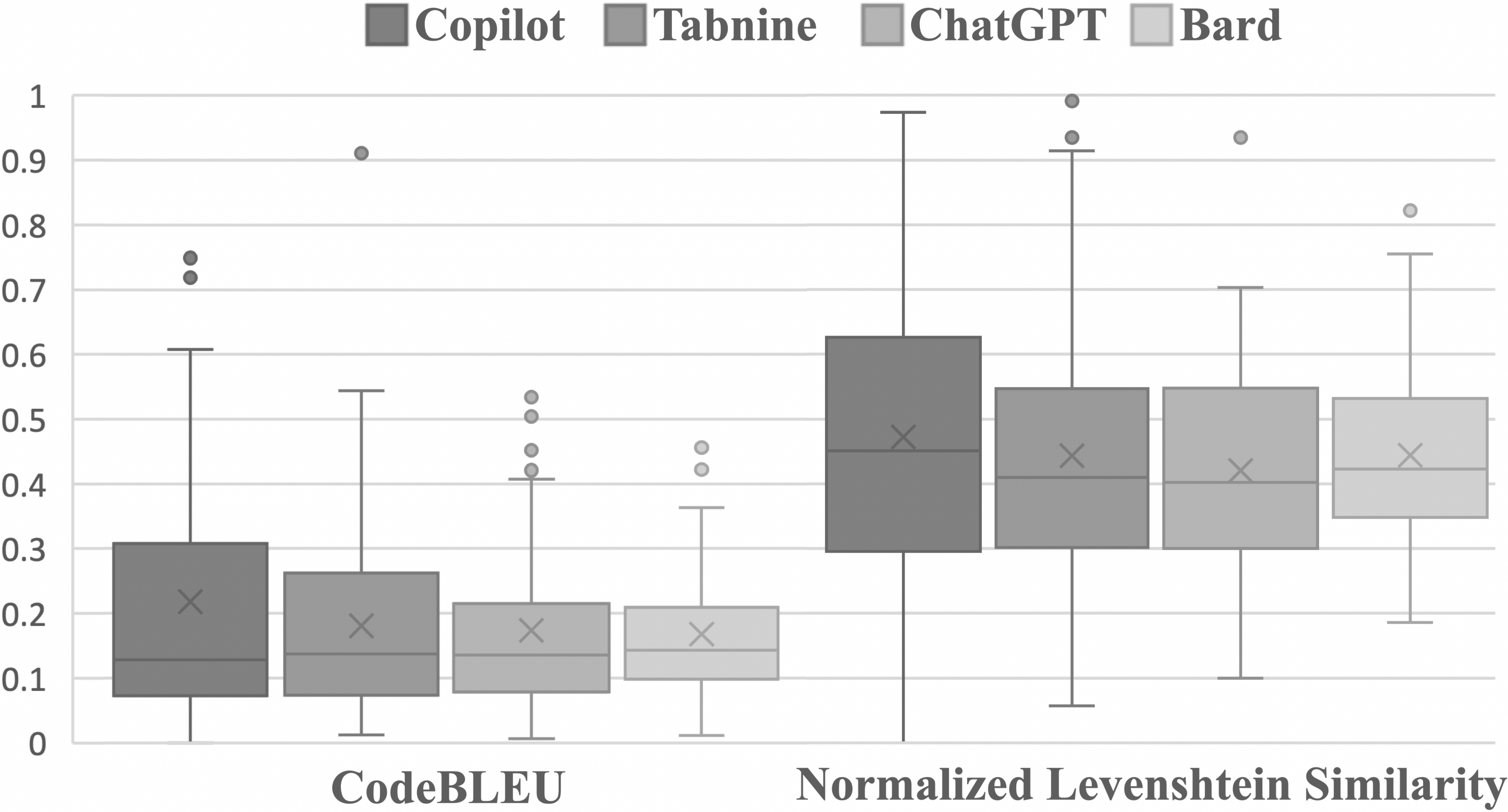}
  \caption{Similarity of incorrect and plausible methods (excluded the correct ones).}
  \label{fig:similarityNotCorrect} \vspace{-0.2cm}
\end{figure}

To derive additional information about the difference between generated and developer code, we compute the CodeBLEU and the normalized Levenshtein similarity, for both the correct methods (Figure~\ref{fig:similarityCorrect}) and the rest of the compilable methods (Figure~\ref{fig:similarityNotCorrect}).  Overall, the CodeBLEU and the Levenshtein similarity analysis help in quantifying and understanding the quality of the generated code, its adherence to developers' styles, and the scope of adjustments needed to make the code correct and aligned with the intended functionality.

In the correct methods, the normalized Levenshtein similarity and the CodeBLEU, with the latter metric reporting lower values than the former one, reveal that several adaptations are needed to adhere to the code and style of the developers. Based on our code inspection, it is often necessary to rename variables, change the order of the statements, reuse code, and refactor methods. This suggests that AI-based code assistants must be improved to adapt to the coding style of the specific project where the code is generated. 

Interestingly Tabnine outperforms all other approaches reaching a median CodeBLEU of 0.528 and demonstrating a better ability to generate code fitting the target project (with statistically significant difference for ChatGPT and Bard, and with no statistical difference for Copilot, according to Mann-Whitney test with significance level 0.05). Notably, Bard and ChatGPT perform slightly worse than Copilot and Tabnine, according to both the normalized Levenshtein similarity and the CodeBLEU. This is probably because Copilot and Tabnine have access to a larger code set, which includes code from a greater variety of developers. This means that they have a better understanding of the different coding styles used by developers.

For the methods that are not correct, the normalized Levenshtein similarity and CodeBLEU are worse, with all approaches requiring a similar, and potentially large, amount of changes to obtain code aligned with one of the developers (no statistically significant differences among values). Although the CodeBLEU and Levenshtein similarity metrics are not a direct measure of effort, a median CodeBLEU close to $0.1$ is indicative of code that is largely misaligned with the developers' code. Stylistic differences may also complicate the correction of the generated code, since its understanding shall require a higher cognitive effort.

When targeting the wrong methods, Tabnine does not outperform anymore the competing techniques. This suggests that Tabnine usually generates either correct code that fits the target project well or wrong code that does not fit well the project. 

\emph{In a nutshell, Tabnine and Copilot demonstrated a better ability to generate code that fits into the target project, with Tabnine performing better than the other tools for the correct methods. Overall, the capability to generate code adhering to the style of the developers requires more research, since results are relatively good for the correct methods and quite poor for the incorrect methods.}

\subsection{Threats to Validity}
In this section, we discuss the main threats affecting our study.

An internal threat to validity concerns the way we used the four compared tools. This threat is mitigated by the kind of well-defined problem we studied, that is, the generation of the body of a method. This scenario results in a  straightforward procedure: we removed the body of the method to be generated, and we asked the assistant to generate the code for us. 

Another source of threat might be the manual and automated analysis of the generated code. To mitigate any implementation threat, we relied on well-established tools for the automated analysis. Concerning the manual analysis, two authors independently assessed the code and then resolved any conflicting classification in a dedicated meeting. This procedure mitigates any risk of misclassification. Moreover, we made all the material publicly available, enabling future independent inspection of our work.

Another internal threat to validity is the possibility of using, as programming tasks, methods already present in the training data used by the compared tools. To mitigate this risk, we controlled the date of creation of the methods used in the evaluation, selecting methods created after the compared services had been released.       
 
We identified the selection of non-representative cases (i.e., methods) for the assessment as a potential external validity threat. To avoid this risk, we carefully defined a method selection process that guarantees the selection of actual methods implemented by developers in well-rated open-source projects, covering a diversity of cases for complexity and dependency from context (see Section~\ref{sec:methodology}). 

A last external validity threat concerns the generalizability of the findings. We do not claim our results generalize beyond the four AI-based code assistants experimented with the generation of Java methods, which already represent a fairly important and representative picture of the state of the art in the area.
\section{Findings} \label{sec:findings}
We summarize here the main findings resulting from our study.

\textbf{No clear winning AI-based code assistant calls for more research about collaboration among multiple AI-based assistants and developers.} On one hand, results obtained for RQ1 show that Copilot is slightly more effective than competing techniques in the generation of code that can be readily integrated into a project. On the other hand, each technique generated correct methods that have not been generated by the other approaches, suggesting developers should not exploit a single AI-based code assistant. This paves the way for more research about the design of solutions for the integration and collaboration among multiple AI-based code assistants and the definition of methodologies to let developers effectively handle suggestions from multiple assistants. 

\textbf{AI-based code assistants may behave better than developers.} Interestingly AI-based code assistants sometimes generated better code than the code generated by the developers, removing code smells, implementing slightly more efficient code, or reducing the complexity of some methods (e.g., through lambda expressions)  (see results for RQs1-3). This suggests the suitability of these assistants for the generation of real-time suggestions to improve code.

\textbf{AI-based code assistants have been quite ineffective, especially with external dependencies.} Although AI-based code assistants are not expected to always produce correct code, results obtained for RQ1 show there is still a significant research gap to be addressed. In fact, a non trivial portion of the generated code is invalid or incorrect (53\% of the methods are either invalid or incorrect for the best performing AI-based code assistant), especially 
when the generated code has to interact with other elements of a project. In fact, the four studied approaches generated only between 4\% and 16\% correct method implementations when dependencies from other classes in the project are present. This result calls for more research, especially to address dependencies from the context.

\textbf{AI-Based code assistants do not well address the style and format of the code.} Teams normally adopt well-defined, and sometimes explicitly enforced (e.g., by tools or during inspection of merge requests) practices in code development. AI-based code assistants should ideally generate code that readily fits into the target project. Results obtained with RQ5 show that this is still largely false. The gap is significant for correct code, and extremely large for the incorrect code, calling for smarter assistants that could consider the context in which code is generated to produce better code. 

\textbf{Checking the correctness of the code generated by \linebreak AI-based assistants uniquely using test cases is not sufficient.} In our experiment, we computed both the set of plausible methods (i.e., methods that pass the available test cases) and the set of correct methods (i.e., methods that are correct based on manual inspection). Results show a significant gap between the number of plausible and correct methods, for all the AI-based assistants. Similarly to what has been already reported for automatic program repair techniques~\cite{Gazzola:Repair:TSE:2019}, the output of code generators requires careful inspection before it can be integrated into an actual project.

   \section{Related Work} \label{sec:related}
Developers are known to exploit IDEs to receive code recommendations. For instance, code completion is reported as one of the most used features in IDEs according to the works by Amann et al.~\cite{Amann:SANER:2016} and Murphy et al.~\cite{Murphy:2006}. So far, several approaches investigated how to recommend code by learning from existing code~\cite{Allamanis:SurveyML:2018}. For instance, Brunch et al.~\cite{Learning:ESECFSE:2009} were among the first to study how to recommend method calls based on features extracted from the code. Other approaches studied how to recommend code based on the nearby code, using n-grams~\cite{Cache:ICSE:2015,Statistical:ESECFSE:2013,Localness:FSE:2014} and code-grammars~\cite{PHOG:ICML:2016}. 

More recently, recommendations produced by AI-based code assistants heavily exploiting large language models trained from huge codebases attracted the attention of researchers and practitioners. Multiple organizations owning data trained and delivered advanced AI-based services, such as GitHub Copilot~\cite{Copilot2023}, Tabnine~\cite{Tabnine2023}, 
ChatGPT~\cite{ChatGPT2023}, and Google Bard~\cite{GoogleBard2023}. 

Since the actual capabilities of these services are yet unclear, several researchers investigated the effectiveness of these services empirically. Vaithilingam et al. conducted a human study about the usability of GitHub Copilot~\cite{Vaithilingam22}. The study analyzes how programmers use and perceive Copilot, revealing that, although Copilot is not always correct, it provides code that is a good starting point to complete programming tasks. 
The study revealed that Copilot may generate code that is not optimal or efficient, and does not always conform to best practices or coding standards. These findings are coherent with our results that show that Copilot, although being the most effective AI-based code assistant, often generates wrong code (67\% of the methods are not correct) and also code that might be inefficient (as discussed in RQ3). While the incorrect code could be a good starting point, the results for the Levenshtein similarity and the CodeBLEU reveal a non-trivial amount of changes needed to reach an implementation consistent with the developers' code. 

Another related study on Copilot is the one by Nguyen et al.~\cite{Nguyen:MSR22}. The study focuses on assessing the correctness and understandability of code generated by Copilot for 33 LeetCode~\cite{LeetCode2023} questions, which represent exemplary problems to get ready for job interview.
The study shows that Copilot can generate 57\% plausible Java methods, while it is less effective with the other languages. In our study, Copilot generated 47\% plausible methods. The slightly lower effectiveness might be due to the real-life nature of the methods that we considered, in comparison to LeetCode questions.

The work by Yetistiren et al.~\cite{Yetistiren:PROMISE:22} investigates Copilot applied to a variety of simple Python problems, such as string manipulation, array manipulation, and sorting, selected from the HumanEval dataset~\cite{HumanEval}. Also in this study, the focus is on identifying the plausible methods generated, and not the correct ones.
Finally, Dakhel et al. considered Copilot applied to algorithmic problems~\cite{Dakhel:JSS:2023}, finding that it can solve fundamental algorithmic problems with high accuracy producing code that is comparable to human solutions.

Our study extends this body of research along many dimensions: we comparatively analyze the effectiveness of four techniques rather than considering Copilot only, we use real methods extracted from open source projects rather than simple exemplary problems, and we consider methods with different levels of complexity and dependency from the context rather than standalone code, delivering new evidence about the capabilities of AI-based code assistants.

The recent work by Yeti{\c{s}}tiren et al.~\cite{yeticstiren2023evaluating} is probably the closest to our work since they also considered multiple tools (GitHub Copilot, Amazon CodeWhisperer, and ChatGPT). 
The study reported the highest success rate for ChatGPT, with 65.2\% plausible solutions generated. This work does not investigate actual code correctness, but only its plausibility, and uses the HumanEval dataset, which is relatively representative of real-world scenarios. 

Finally, the CrystalBLEU metric has been recently proposed as an alternative metric to the CodeBLEU metric~\cite{Eghbali:CrystalBLEU:ASE:2022}. In the future, we plan to study code similarity also according to this metric.

\section{Conclusions} \label{sec:conclusion}
AI-based code assistants are gaining popularity thanks to the quick advances in deep learning and large language models. Recent studies suggest that these tools may change the way developers implement code, ultimately speeding up the development process. 
This paper contributes to expanding the knowledge in this area by reporting an empirical study based on methods extracted from open-source projects, whose selection has been controlled along many dimensions, including the relevance and diversity of the projects, the complexity and degree of dependencies from the context, and the lack of overlap with the training set of the compared tools. 
Further, our study investigates many dimensions including the manually assessed correctness of the generated code, comparison of four alternative approaches, and analysis of several characteristics, including complexity, efficiency, and size. Results highlight the complementarities among the compared AI-based code assistants and result in findings that may influence future research. 

Future work concerns with exploiting this evidence to elaborate appropriate protocols to work with the code produced by AI-based code-assistants, in addition to extending our study to other languages and aspects, \begin{change} such as code repair.\end{change}

\begin{acks}
This work has been partially supported by the Engineered MachinE Learning-intensive IoT systems (EMELIOT) national research project (PRIN 2020 program Contract 2020W3A5FY) and the MUR under the grant "Dipartimenti di Eccellenza 2023-2027" of the Department of Informatics, Systems and Communication of the University of Milano-Bicocca, Italy.
\end{acks}
\balance
\bibliographystyle{ACM-Reference-Format}

\begin{thebibliography}{34}



\ifx \showCODEN    \undefined \def \showCODEN     #1{\unskip}     \fi
\ifx \showDOI      \undefined \def \showDOI       #1{#1}\fi
\ifx \showISBNx    \undefined \def \showISBNx     #1{\unskip}     \fi
\ifx \showISBNxiii \undefined \def \showISBNxiii  #1{\unskip}     \fi
\ifx \showISSN     \undefined \def \showISSN      #1{\unskip}     \fi
\ifx \showLCCN     \undefined \def \showLCCN      #1{\unskip}     \fi
\ifx \shownote     \undefined \def \shownote      #1{#1}          \fi
\ifx \showarticletitle \undefined \def \showarticletitle #1{#1}   \fi
\ifx \showURL      \undefined \def \showURL       {\relax}        \fi
\providecommand\bibfield[2]{#2}
\providecommand\bibinfo[2]{#2}
\providecommand\natexlab[1]{#1}
\providecommand\showeprint[2][]{arXiv:#2}

\bibitem[\protect\citeauthoryear{Allamanis, Barr, Devanbu, and
  Sutton}{Allamanis et~al\mbox{.}}{2018}]{Allamanis:SurveyML:2018}
\bibfield{author}{\bibinfo{person}{Miltiadis Allamanis},
  \bibinfo{person}{Earl~T. Barr}, \bibinfo{person}{Premkumar Devanbu}, {and}
  \bibinfo{person}{Charles Sutton}.} \bibinfo{year}{2018}\natexlab{}.
\newblock \showarticletitle{A Survey of Machine Learning for Big Code and
  Naturalness}.
\newblock \bibinfo{journal}{\emph{Comput. Surveys}} \bibinfo{volume}{51},
  \bibinfo{number}{4} (\bibinfo{year}{2018}).
\newblock
\urldef\tempurl \url{https://doi.org/10.1145/3212695}
\showDOI{\tempurl}


\bibitem[\protect\citeauthoryear{Amann, Proksch, Nadi, and Mezini}{Amann
  et~al\mbox{.}}{2016}]{Amann:SANER:2016}
\bibfield{author}{\bibinfo{person}{Sven Amann}, \bibinfo{person}{Sebastian
  Proksch}, \bibinfo{person}{Sarah Nadi}, {and} \bibinfo{person}{Mira Mezini}.}
  \bibinfo{year}{2016}\natexlab{}.
\newblock \showarticletitle{A Study of Visual Studio Usage in Practice}. In
  \bibinfo{booktitle}{\emph{Proceedings of the International Conference on
  Software Analysis, Evolution, and Reengineering (SANER)}}.
\newblock
\urldef\tempurl \url{https://doi.org/10.1109/SANER.2016.39}
\showDOI{\tempurl}


\bibitem[\protect\citeauthoryear{Apache}{Apache}{2023}]{SureFirePlugin2023}
\bibfield{author}{\bibinfo{person}{Apache}.} \bibinfo{year}{2023}\natexlab{}.
\newblock \bibinfo{title}{Maven Surefire Plugin}.
\newblock
\newblock
\newblock
\shownote{\url{https://maven.apache.org/surefire/maven-surefire-plugin/}.}


\bibitem[\protect\citeauthoryear{Bielik, Raychev, and Vechev}{Bielik
  et~al\mbox{.}}{2016}]{PHOG:ICML:2016}
\bibfield{author}{\bibinfo{person}{Pavol Bielik}, \bibinfo{person}{Veselin
  Raychev}, {and} \bibinfo{person}{Martin Vechev}.}
  \bibinfo{year}{2016}\natexlab{}.
\newblock \showarticletitle{PHOG: Probabilistic Model for Code}. In
  \bibinfo{booktitle}{\emph{Proceedings of the International Conference on
  Machine Learning (ICML)}}.
\newblock


\bibitem[\protect\citeauthoryear{Bruch, Monperrus, and Mezini}{Bruch
  et~al\mbox{.}}{2009}]{Learning:ESECFSE:2009}
\bibfield{author}{\bibinfo{person}{Marcel Bruch}, \bibinfo{person}{Martin
  Monperrus}, {and} \bibinfo{person}{Mira Mezini}.}
  \bibinfo{year}{2009}\natexlab{}.
\newblock \showarticletitle{Learning from Examples to Improve Code Completion
  Systems}. In \bibinfo{booktitle}{\emph{Proceedings of the Joint Meeting of
  the European Software Engineering Conference and the ACM SIGSOFT Symposium on
  the Foundations of Software Engineering (ESEC/FSE)}}.
\newblock
\urldef\tempurl \url{https://doi.org/10.1145/1595696.1595728}
\showDOI{\tempurl}


\bibitem[\protect\citeauthoryear{Eghbali and Pradel}{Eghbali and
  Pradel}{2022}]{Eghbali:CrystalBLEU:ASE:2022}
\bibfield{author}{\bibinfo{person}{Aryaz Eghbali} {and}
  \bibinfo{person}{Michael Pradel}.} \bibinfo{year}{2022}\natexlab{}.
\newblock \showarticletitle{CrystalBLEU: Precisely and Efficiently Measuring
  the Similarity of Code}. In \bibinfo{booktitle}{\emph{Proceedings of the
  International Conference on Automated Software Engineering (ASE)}}.
\newblock
\urldef\tempurl \url{https://doi.org/10.1145/3551349.3556903}
\showDOI{\tempurl}


\bibitem[\protect\citeauthoryear{Fagadau, Mariani, Micucci, and
  Riganelli}{Fagadau et~al\mbox{.}}{2024}]{Fagadau:Prompt:ICPC:2024}
\bibfield{author}{\bibinfo{person}{Ionut~Daniel Fagadau},
  \bibinfo{person}{Leonardo Mariani}, \bibinfo{person}{Daniela Micucci}, {and}
  \bibinfo{person}{Oliviero Riganelli}.} \bibinfo{year}{2024}\natexlab{}.
\newblock \showarticletitle{Analyzing Prompt Influence on Automated Method
  Generation: An Empirical Study with Copilot}. In
  \bibinfo{booktitle}{\emph{Proceedings of the International Conference on
  Program Comprehension (ICPC)}}.
\newblock
\urldef\tempurl \url{https://doi.org/10.1145/3643916.3644409}
\showDOI{\tempurl}


\bibitem[\protect\citeauthoryear{Forge}{Forge}{2023}]{DependencyFinder}
\bibfield{author}{\bibinfo{person}{Source Forge}.}
  \bibinfo{year}{2023}\natexlab{}.
\newblock \bibinfo{title}{DependencyFinder}.
\newblock
\newblock
\newblock
\shownote{\url{https://scitools.com/}.}


\bibitem[\protect\citeauthoryear{Franks, Tu, Devanbu, and Hellendoorn}{Franks
  et~al\mbox{.}}{2015}]{Cache:ICSE:2015}
\bibfield{author}{\bibinfo{person}{Christine Franks}, \bibinfo{person}{Zhaopeng
  Tu}, \bibinfo{person}{Premkumar~T. Devanbu}, {and} \bibinfo{person}{Vincent
  Hellendoorn}.} \bibinfo{year}{2015}\natexlab{}.
\newblock \showarticletitle{Cacheca: A cache language model based code
  suggestion tool}. In \bibinfo{booktitle}{\emph{Proceedings of the
  International Conference on Software Engineering (ICSE)}}.
\newblock
\urldef\tempurl \url{https://doi.org/doi: 10.1109/ICSE.2015.228}
\showDOI{\tempurl}


\bibitem[\protect\citeauthoryear{Gazzola, Micucci, and Mariani}{Gazzola
  et~al\mbox{.}}{2019}]{Gazzola:Repair:TSE:2019}
\bibfield{author}{\bibinfo{person}{Luca Gazzola}, \bibinfo{person}{Daniela
  Micucci}, {and} \bibinfo{person}{Leonardo Mariani}.}
  \bibinfo{year}{2019}\natexlab{}.
\newblock \showarticletitle{Automatic Software Repair: A Survey}.
\newblock \bibinfo{journal}{\emph{IEEE Transactions on Software Engineering}}
  \bibinfo{volume}{45}, \bibinfo{number}{1} (\bibinfo{year}{2019}),
  \bibinfo{pages}{34--67}.
\newblock


\bibitem[\protect\citeauthoryear{GitHub}{GitHub}{2023}]{Copilot2023}
\bibfield{author}{\bibinfo{person}{GitHub}.} \bibinfo{year}{2023}\natexlab{}.
\newblock \bibinfo{title}{Copilot}.
\newblock
\newblock
\newblock
\shownote{\url{https://github.com/features/copilot}.}


\bibitem[\protect\citeauthoryear{Google}{Google}{2023}]{GoogleBard2023}
\bibfield{author}{\bibinfo{person}{Google}.} \bibinfo{year}{2023}\natexlab{}.
\newblock \bibinfo{title}{Bard}.
\newblock
\newblock
\newblock
\shownote{\url{https://bard.google.com}.}


\bibitem[\protect\citeauthoryear{LeetCode}{LeetCode}{2023}]{LeetCode2023}
\bibfield{author}{\bibinfo{person}{LeetCode}.} \bibinfo{year}{2023}\natexlab{}.
\newblock \bibinfo{title}{LeetCode}.
\newblock
\newblock
\newblock
\shownote{\url{https://leetcode.com}.}


\bibitem[\protect\citeauthoryear{Little and Miller}{Little and Miller}{2007}]{little2007keyword}
\bibfield{author}{\bibinfo{person}{Greg Little} {and} \bibinfo{person}{Robert~C
  Miller}.} \bibinfo{year}{2007}\natexlab{}.
\newblock \showarticletitle{Keyword programming in Java}. In
  \bibinfo{booktitle}{\emph{Proceedings of the IEEE/ACM International
  Conference on Automated Software Engineering (ICSE)}}.
\newblock
\urldef\tempurl \url{https://doi.org/10.1145/1321631.1321646}
\showDOI{\tempurl}


\bibitem[\protect\citeauthoryear{McCabe}{McCabe}{1976}]{McCabe76}
\bibfield{author}{\bibinfo{person}{Thomas~J. McCabe}.}
  \bibinfo{year}{1976}\natexlab{}.
\newblock \showarticletitle{A Complexity Measure}.
\newblock \bibinfo{journal}{\emph{IEEE Transactions on Software Engineering}}
  \bibinfo{volume}{SE-2}, \bibinfo{number}{4} (\bibinfo{year}{1976}),
  \bibinfo{pages}{308--320}.
\newblock
\urldef\tempurl \url{https://doi.org/doi: 10.1109/TSE.1976.233837}
\showDOI{\tempurl}


\bibitem[\protect\citeauthoryear{Microsoft}{Microsoft}{2020}]{CodeXGLUE}
\bibfield{author}{\bibinfo{person}{Microsoft}.}
  \bibinfo{year}{2020}\natexlab{}.
\newblock \bibinfo{title}{CodeXGLUE}.
\newblock
\newblock
\newblock
\shownote{\url{https://github.com/microsoft/CodeXGLUE/tree/main/Code-Code/code-to-code-trans/evaluator}.}


\bibitem[\protect\citeauthoryear{Microsoft}{Microsoft}{2023}]{VisualStudioCode2023}
\bibfield{author}{\bibinfo{person}{Microsoft}.}
  \bibinfo{year}{2023}\natexlab{}.
\newblock \bibinfo{title}{Visual Studio Code}.
\newblock
\newblock
\newblock
\shownote{\url{https://code.visualstudio.com}.}


\bibitem[\protect\citeauthoryear{{Moradi Dakhel}, Majdinasab, Nikanjam, Khomh,
  Desmarais, and Jiang}{{Moradi Dakhel} et~al\mbox{.}}{2023}]{Dakhel:JSS:2023}
\bibfield{author}{\bibinfo{person}{Arghavan {Moradi Dakhel}},
  \bibinfo{person}{Vahid Majdinasab}, \bibinfo{person}{Amin Nikanjam},
  \bibinfo{person}{Foutse Khomh}, \bibinfo{person}{Michel~C. Desmarais}, {and}
  \bibinfo{person}{Zhen Ming~(Jack) Jiang}.} \bibinfo{year}{2023}\natexlab{}.
\newblock \showarticletitle{GitHub Copilot AI pair programmer: Asset or
  Liability?}
\newblock \bibinfo{journal}{\emph{Journal of Systems and Software}}
  \bibinfo{volume}{203} (\bibinfo{year}{2023}).
\newblock
\urldef\tempurl \url{https://doi.org/10.1016/j.jss.2023.111734}
\showDOI{\tempurl}


\bibitem[\protect\citeauthoryear{Murphy, Kersten, and Findlater}{Murphy
  et~al\mbox{.}}{2006}]{Murphy:2006}
\bibfield{author}{\bibinfo{person}{Gail~C. Murphy}, \bibinfo{person}{Mik
  Kersten}, {and} \bibinfo{person}{Leah Findlater}.}
  \bibinfo{year}{2006}\natexlab{}.
\newblock \showarticletitle{How are Java software developers using the Eclipse
  IDE?}
\newblock \bibinfo{journal}{\emph{IEEE Software}} \bibinfo{volume}{23},
  \bibinfo{number}{4} (\bibinfo{year}{2006}), \bibinfo{pages}{76--83}.
\newblock
\urldef\tempurl \url{https://doi.org/10.1109/MS.2006.105}
\showDOI{\tempurl}


\bibitem[\protect\citeauthoryear{Nguyen and Nadi}{Nguyen and Nadi}{2022}]{Nguyen:MSR22}
\bibfield{author}{\bibinfo{person}{Nhan Nguyen} {and} \bibinfo{person}{Sarah
  Nadi}.} \bibinfo{year}{2022}\natexlab{}.
\newblock \showarticletitle{An Empirical Evaluation of GitHub Copilot's Code
  Suggestions}. In \bibinfo{booktitle}{\emph{Proceedings of the International
  Conference on Mining Software Repositories (MSR)}}.
\newblock
\urldef\tempurl \url{https://doi.org/10.1145/3524842.3528470}
\showDOI{\tempurl}


\bibitem[\protect\citeauthoryear{Nguyen, Nguyen, Nguyen, and Nguyen}{Nguyen
  et~al\mbox{.}}{2013}]{Statistical:ESECFSE:2013}
\bibfield{author}{\bibinfo{person}{Tung~Thanh Nguyen},
  \bibinfo{person}{Anh~Tuan Nguyen}, \bibinfo{person}{Hoan~Anh Nguyen}, {and}
  \bibinfo{person}{Tien~N. Nguyen}.} \bibinfo{year}{2013}\natexlab{}.
\newblock \showarticletitle{A statistical semantic language model for source
  code}. In \bibinfo{booktitle}{\emph{Proceedings of the Joint Meeting of the
  European Software Engineering Conference and the Symposium on the Foundations
  of Software Engineering (ESEC/FSE)}}.
\newblock
\urldef\tempurl \url{https://doi.org/10.1145/2491411.2491458}
\showDOI{\tempurl}


\bibitem[\protect\citeauthoryear{NumPy}{NumPy}{2022}]{linspace}
\bibfield{author}{\bibinfo{person}{NumPy}.} \bibinfo{year}{2022}\natexlab{}.
\newblock \bibinfo{title}{linspace}.
\newblock
\newblock
\newblock
\shownote{\url{https://numpy.org/doc/stable/reference/generated/numpy.linspace.html}.}


\bibitem[\protect\citeauthoryear{OpenAI}{OpenAI}{2023a}]{ChatGPT2023}
\bibfield{author}{\bibinfo{person}{OpenAI}.} \bibinfo{year}{2023}\natexlab{a}.
\newblock \bibinfo{title}{ChatGPT}.
\newblock
\newblock
\newblock
\shownote{\url{https://openai.com/chatgpt}.}


\bibitem[\protect\citeauthoryear{OpenAI}{OpenAI}{2023b}]{HumanEval}
\bibfield{author}{\bibinfo{person}{OpenAI}.} \bibinfo{year}{2023}\natexlab{b}.
\newblock \bibinfo{title}{HumalEval}.
\newblock
\newblock
\newblock
\shownote{\url{https://github.com/openai/human-eval}.}


\bibitem[\protect\citeauthoryear{Proksch, Amann, and Nadi}{Proksch
  et~al\mbox{.}}{2018}]{proksch2018enriched}
\bibfield{author}{\bibinfo{person}{Sebastian Proksch}, \bibinfo{person}{Sven
  Amann}, {and} \bibinfo{person}{Sarah Nadi}.} \bibinfo{year}{2018}\natexlab{}.
\newblock \showarticletitle{Enriched event streams: a general dataset for
  empirical studies on in-IDE activities of software developers}. In
  \bibinfo{booktitle}{\emph{Proceedings of the International Conference on
  Mining Software Repositories (MSR)}}.
\newblock
\urldef\tempurl \url{https://doi.org/10.1145/3196398.3196400}
\showDOI{\tempurl}


\bibitem[\protect\citeauthoryear{PyPI}{PyPI}{2023}]{Levenshteinsw}
\bibfield{author}{\bibinfo{person}{PyPI}.} \bibinfo{year}{2023}\natexlab{}.
\newblock \bibinfo{title}{python-Levenshtein 0.21.1}.
\newblock
\newblock
\newblock
\shownote{\url{https://pypi.org/project/python-Levenshtein/}.}


\bibitem[\protect\citeauthoryear{Raychev, Vechev, and Yahav}{Raychev
  et~al\mbox{.}}{2014}]{raychev2014code}
\bibfield{author}{\bibinfo{person}{Veselin Raychev}, \bibinfo{person}{Martin
  Vechev}, {and} \bibinfo{person}{Eran Yahav}.}
  \bibinfo{year}{2014}\natexlab{}.
\newblock \showarticletitle{Code completion with statistical language models}.
\newblock \bibinfo{journal}{\emph{SIGPLAN Not.}} \bibinfo{volume}{49},
  \bibinfo{number}{6} (\bibinfo{year}{2014}).
\newblock
\urldef\tempurl \url{https://doi.org/10.1145/2666356.2594321}
\showDOI{\tempurl}


\bibitem[\protect\citeauthoryear{Ren, Guo, Lu, Zhou, Liu, Tang, Sundaresan,
  Zhou, Blanco, and Ma}{Ren et~al\mbox{.}}{2020}]{CodeBLEU}
\bibfield{author}{\bibinfo{person}{Shuo Ren}, \bibinfo{person}{Daya Guo},
  \bibinfo{person}{Shuai Lu}, \bibinfo{person}{Long Zhou},
  \bibinfo{person}{Shujie Liu}, \bibinfo{person}{Duyu Tang},
  \bibinfo{person}{Neel Sundaresan}, \bibinfo{person}{Ming Zhou},
  \bibinfo{person}{Ambrosio Blanco}, {and} \bibinfo{person}{Shuai Ma}.}
  \bibinfo{year}{2020}\natexlab{}.
\newblock \bibinfo{title}{CodeBLEU: a Method for Automatic Evaluation of Code
  Synthesis}.
\newblock
\newblock
\showeprint[arxiv]{2009.10297}


\bibitem[\protect\citeauthoryear{SciTppls}{SciTppls}{2023}]{understand}
\bibfield{author}{\bibinfo{person}{SciTppls}.} \bibinfo{year}{2023}\natexlab{}.
\newblock \bibinfo{title}{DependencyFinder}.
\newblock
\newblock
\newblock
\shownote{\url{https://depfind.sourceforge.io}.}


\bibitem[\protect\citeauthoryear{Tabnine}{Tabnine}{2023}]{Tabnine2023}
\bibfield{author}{\bibinfo{person}{Tabnine}.} \bibinfo{year}{2023}\natexlab{}.
\newblock \bibinfo{title}{Tabnine}.
\newblock
\newblock
\newblock
\shownote{\url{https://www.tabnine.com}.}


\bibitem[\protect\citeauthoryear{Tu, Su, and Devanbu}{Tu et~al\mbox{.}}{2014}]{Localness:FSE:2014}
\bibfield{author}{\bibinfo{person}{Zhaopeng Tu}, \bibinfo{person}{Zhendong Su},
  {and} \bibinfo{person}{Premkumar Devanbu}.} \bibinfo{year}{2014}\natexlab{}.
\newblock \showarticletitle{On the localness of software}. In
  \bibinfo{booktitle}{\emph{Proceedings of the International Symposium on
  Foundations of Software Engineering (FSE)}}.
\newblock
\urldef\tempurl \url{https://doi.org/10.1145/2635868.2635875}
\showDOI{\tempurl}


\bibitem[\protect\citeauthoryear{Vaithilingam, Zhang, and
  Glassman}{Vaithilingam et~al\mbox{.}}{2022}]{Vaithilingam22}
\bibfield{author}{\bibinfo{person}{Priyan Vaithilingam},
  \bibinfo{person}{Tianyi Zhang}, {and} \bibinfo{person}{Elena~L. Glassman}.}
  \bibinfo{year}{2022}\natexlab{}.
\newblock \showarticletitle{Expectation vs. Experience: Evaluating the
  Usability of Code Generation Tools Powered by Large Language Models}. In
  \bibinfo{booktitle}{\emph{Extended Abstracts of the CHI Conference on Human
  Factors in Computing Systems (CHI EA)}}.
\newblock
\urldef\tempurl \url{https://doi.org/10.1145/3491101.3519665}
\showDOI{\tempurl}


\bibitem[\protect\citeauthoryear{Yeti{\c{s}}tiren, {\"O}zsoy, Ayerdem, and
  T{\"u}z{\"u}n}{Yeti{\c{s}}tiren et~al\mbox{.}}{2023}]{yeticstiren2023evaluating}
\bibfield{author}{\bibinfo{person}{Burak Yeti{\c{s}}tiren},
  \bibinfo{person}{I{\c{s}}{\i}k {\"O}zsoy}, \bibinfo{person}{Miray Ayerdem},
  {and} \bibinfo{person}{Eray T{\"u}z{\"u}n}.} \bibinfo{year}{2023}\natexlab{}.
\newblock \bibinfo{title}{Evaluating the Code Quality of AI-Assisted Code
  Generation Tools: An Empirical Study on GitHub Copilot, Amazon CodeWhisperer,
  and ChatGPT}.
\newblock
\newblock
\showeprint[arxiv]{2304.10778}


\bibitem[\protect\citeauthoryear{Yetistiren, Ozsoy, and Tuzun}{Yetistiren
  et~al\mbox{.}}{2022}]{Yetistiren:PROMISE:22}
\bibfield{author}{\bibinfo{person}{Burak Yetistiren}, \bibinfo{person}{Isik
  Ozsoy}, {and} \bibinfo{person}{Eray Tuzun}.} \bibinfo{year}{2022}\natexlab{}.
\newblock \showarticletitle{Assessing the Quality of GitHub Copilot’s Code
  Generation}. In \bibinfo{booktitle}{\emph{Proceedings of the International
  Conference on Predictive Models and Data Analytics in Software Engineering
  (PROMISE)}}.
\newblock
\urldef\tempurl \url{https://doi.org/10.1145/3558489.3559072}
\showDOI{\tempurl}


\end{thebibliography}

\end{document}